\documentclass[twocolumn,showpacs,preprintnumbers,amsmath,amssymb,10pt,a4paper]{revtex4}
\usepackage{geometry}
\usepackage{graphicx}% Include figure files
\usepackage{dcolumn}% Align table columns on decimal point
\usepackage{bm}% bold math
\usepackage[english]{babel}
\usepackage{color}

\newcommand{\be}{\begin{equation}}
\newcommand{\ee}{\end{equation}}
\newcommand{\BE}{\begin{eqnarray}}
\newcommand{\EE}{\end{eqnarray}}

\newcommand{\dv}[2]{\frac{{d} #1}{{d} #2} }
\newcommand{\bs}[1]{\boldsymbol{#1}}

\newcommand{\id}{{\rm 1\!\!I}}

\newcommand{\bx}{\ensuremath{\mathbf{x}}}

\newcommand{\bF}{\ensuremath{\mathbf{F}}}

\newcommand{\avg}[1]{\left\langle{#1}\right\rangle}

\newcommand{\bxi}{{\mbox{\boldmath $\xi$}}}

\newcommand{\bchi}{{\mbox{\boldmath $\chi$}}}
\newcommand{\bnu}{{\mbox{\boldmath $\nu$}}}
\newcommand{\bzeta}{{\mbox{\boldmath $\zeta$}}}

\begin{document}
%\preprint{}
\title{Demographic noise and piecewise deterministic {M}arkov processes}

\author{John Realpe-Gomez}\email{john.realpegomez@manchester.ac.uk}
\affiliation{Theoretical Physics, School of Physics and Astronomy, 
The University of Manchester, Manchester M13 9PL, United Kingdom}

\author{Tobias Galla}\email{tobias.galla@manchester.ac.uk}
\affiliation{Theoretical Physics, School of Physics and Astronomy, 
The University of Manchester, Manchester M13 9PL, United Kingdom}

\author{Alan J. McKane}\email{alan.mckane@manchester.ac.uk}
\affiliation{Theoretical Physics, School of Physics and Astronomy, 
The University of Manchester, Manchester M13 9PL, United Kingdom}

\date{\today}

\begin{abstract}
We explore a class of hybrid (piecewise deterministic) systems characterized by a large number of individuals inhabiting an environment whose state is described by a set of continuous variables. We use analytical and numerical methods from nonequilibrium statistical mechanics to study the influence that intrinsic noise has on the qualitative behavior of the system. We discuss the application of these concepts to the case of semi-arid ecosystems. Using a system-size expansion we calculate the power spectrum of the fluctuations in the system. This predicts the existence of noise-induced oscillations. 
\end{abstract}


\maketitle

%%%%%%%%%%%%%%%%%%%%%%%%%%%%%%%%%%%%%%%%%%%%%%%%%%%%%%%%%%%%%%%%%%%%%%%%%%%%%%
%%%%%%%%%%%%%%%%%%%%%%%%%%%%%%%%%%%%%%%%%%%%%%%%%%%%%%%%%%%%%%%%%%%%%%%%%%%%%%

%\nocite{*}
\section{Introduction}
\label{intro}

When modelling complex systems, we often have to decide between two contrasting approaches: whether we focus on the behavior of the individuals that constitute the system, or on the aggregate behavior in the large. Furthermore, there is usually a trade-off between realism and tractability: while individual based models (IBMs) are versatile enough to incorporate many details of the system of interest, they can soon become computationally unmanageable if too many details are included. Moreover, with too much detail we may end up sacrificing insight into the way that the model works. Similarly, while aggregate models, usually based on differential equations for a continuous density of individuals, can be far more tractable and provide powerful insights  through the use of analytical tools, they are susceptible to missing relevant details that can affect even the {\it qualitative} predictions of the model~\cite{McKane-PRL2005,Boland-2008,Boland-2009,Rogers-2012,Biancalani-2011, Grima-Nature, Goldenfeld-2011,Scott-2011,Biancalani-2010,Goldenfeld-2009,Lugo-2008} (see also \cite{McKane-TREE} for a recent general review). It seems, then, that an intermediate approach is desirable in order to find a good compromise between realism, efficiency, and insight.
 
Stochastic hybrid models~\cite{Cassandra-2006, Hespanha-2006, Pola-2003, Sumpter-2006, Davis-1993, Davis-1984} are a general class of models that can be very useful in exploiting the benefits of both of these approaches. Of particular interest to us here is the subclass of so-called piecewise deterministic {M}arkov processes (PDMP)~\cite{Davis-1993,Davis-1984}, which has recently gained increased attention in the natural sciences~\cite{Faggionato-2009, Faggionato-2010, Zeiser-2008, Zeiser-2010, Crudu-press}. A PDMP models a system characterized by both discrete and continuous variables, where the former follow a stochastic process while the latter are governed by a deterministic (differential) equation. These two dynamics are coupled as the dynamical law for each depends on the current state of both. Motivated by applications to ecology, here we assume that the discrete variables refer to the number of individuals (e.g. plants, animals, etc.) that are immersed in an environment whose state is in turn characterized by the continuous variables (e.g. amount of water, temperature, etc.).

Recently, Faggionato et al.~\cite{Faggionato-2009, Faggionato-2010} have initiated a study of PDMP by using tools from nonequilibrium statistical physics. In particular, these authors have obtained some formal results concerning the deterministic asymptotic behavior of the system, as well as the fluctuations about it, when increasing the rate of the stochastic transitions of the discrete variables. Here, we are also interested in the nonequilibrium statistical mechanics of PDMP; however, we will be concerned with the {\it qualitative} differences in behavior from the deterministic asymptotic limit, induced by the fluctuations, especially the existence of noise-induced quasi-oscillations in a regime where the deterministic approximation predicts only a stable fixed point. Furthermore, in contrast to the work of Faggionato et al., we do not restrict the discrete variables to a finite set, but allow them to take on any non-negative integer value. Indeed, the mesoscopic limit we consider assumes a relatively large number of individuals, in the spirit of the system-size expansion developed by van Kampen~\cite{vanKampen-book}. 

The techniques we use are general and can be applied to any system that could be modelled in a piece-wise deterministic fashion. An example could be an environment characterized by a set of continuous variables which would evolve deterministically were it not for the influence of a finite number of individuals that inhabit it. Here we will consider an application where the environmental variables are water in an ecosystem which contains a population of plants. Traditionally, such systems have been modelled by a fully deterministic dynamics of continuous densities. However, following the discussion above, it would seem more natural to model plants as discrete entities, rather than to define a density of plants. It is also well known that spatial interactions play a relevant role in these systems, as the emergence of spatial patterns are common \cite{Meron-2011,Rietkerk-2010,Rietkerk-2008,Meron-2004,Rietkerk-2002,Meron-2001}. In this present paper, however, we will only consider non-spatial models, implicitly implying that their area is sufficiently large that we can neglect any spatial structure altogether. A large region can of course contain a relatively large number of individuals, though still finite. As we will see below, demographic noise can still be relevant in such a system, and induce behavior qualitatively different from that described by deterministic approximations. For smaller numbers of plants these effects are expected to be even stronger, emphasizing the importance of stochastic modelling in this field. 

The paper is organized as follows. In Section \ref{s:gm} we define the general framework and introduce the model for semi-arid ecosystems which will serve as an application of the ideas and techniques which we develop. In Section \ref{s:gme} we show how to derive the master equation that governs the evolution of the system and in Section \ref{s:ad} we discuss the idea behind the system-size expansion which we use in the analysis. In Section \ref{s:dl} we obtain the leading term in this approximation which leads to the non-spatial version of deterministic equations extensively studied in the literature~\cite{Rietkerk-2010,Rietkerk-2008,Rietkerk-2002}. The next-to-leading term in the expansion, which characterizes the fluctuations in the system, is described in Section \ref{s:f}. Finally, in Section \ref{s:c} we present our conclusions. Most of the technical details are collected in the Appendixes.

%%%%%%%%%%%%%%%%%%%%%%%%%
\section{Model definitions}\label{s:gm}
%%%%%%%%%%%%%%%%%%%%%%%%%
\subsection{General setup}
The simplest class of hybrid systems have states that are described by a pair of variables $(n, x)$, where $n$ is discrete and $x$ is continuous. The former would typically be the number of individuals (e.g. plants) in the system, while the latter would refer to the state of the environment (e.g. amount of water) which these individuals inhabit. The dynamics of the continuous variables $x$ is deterministic if conditioned on the discrete variables, $n$. On the other hand, the discrete variables follow a stochastic process whose transition probabilities depend on the continuous variables $x$. In other words, the continuous variables are governed by a deterministic differential equation of the form
\begin{equation}\label{e:general}
\dv{x}{t} = F(n, x) . 
\end{equation}
By contrast, the discrete variables follow some stochastic transition rules that describe the processes which individuals, denoted by $P$, undergo. For instance, in the specific case of a birth-death process, these have the form
\begin{subequations}\label{e:general_rules}
\begin{align}
P &\xrightarrow{\Gamma_b(x)} 2 P,\\
P &\xrightarrow{\Gamma_d(x)} \emptyset .
\end{align}
\end{subequations}
Here $\Gamma_b$ and $\Gamma_d$ are the birth and death rates, respectively, and can depend on the environmental variables $x$. In principle, the transition rules (\ref{e:general_rules}) can describe any other kind of processes such as migration or growth. However, in this paper we will focus exclusively on simple birth-death processes. This constitutes an instance of a piecewise deterministic Markov process (PDMP) \cite{Davis-1984, Davis-1993, Faggionato-2009, Faggionato-2010}; the reason for the adoption of this name will become clear later.

If the system is in state $(n, x)$ the transition rates are then given by
\begin{subequations}\label{e:general_transitions}
\begin{align}
{T}_b(n+1|n;\, x) &= n\, \Gamma_b(x), \\
{T}_d(n-1|n;\, x) &= n\, \Gamma_d(x). 
\end{align}
\end{subequations}

All of this can be generalized to a system with $D$ discrete variables and $C$ continuous variables by introducing the state variables $\mathbf{n}=(n_1,n_2,\ldots,n_D)$ and $\mathbf{x}=(x_1,x_2,\ldots,x_C)$. Having specified the state variables and the transition rates, we can now go on to write down the master equation that governs the evolution of the probability function of the system. However, before doing so, it is useful to give a specific example of a PDMP in order to make these concepts more concrete. 

%%%%%%%%%%%%%%%%%%%%%%%%%
\subsection{Example: A semi-arid ecosystem}\label{s:ex}
%%%%%%%%%%%%%%%%%%%%%%%%%
As an example, we will consider a non-spatial piecewise deterministic model of semi-arid ecosystems, defined in terms of three variables corresponding to, respectively, the densities of surface water, $\sigma$, soil water, $\omega$, and number of plants, $n$. A spatial version of this model has been well-studied in the ecological literature \cite{Rietkerk-2001,Rietkerk-2002,Rietkerk-2008,Rietkerk-2010}, but in the case where the number of plants is effectively infinite. In the model, rainwater falls onto the surface of the land, and then infiltrates into the soil where it is taken up by plants. Although rainfall in a semi-arid environment can vary drastically and unpredictably over time scales short in comparison with the birth-death dynamics of plants, here we focus on the {\it average} amount of rainfall over a relatively long period of time. This is a simplifying assumption that allows us to separate the influence of intrinsic demographic fluctuations from extrinsic environmental noise (see e.g. \cite{D'Odorico-Book, Kettler-2009, Mara-2008, Mara-2007} for investigations on the latter). Indeed, this simplification has also been used by ecologists \cite{Rietkerk-2001,Rietkerk-2002,Rietkerk-2008,Rietkerk-2010} in order to avoid further complications related to the infiltration of water under high water densities (see e.g. \citep{Porporato-Book}). In comparison to these investigations, the only sacrifice of realism in the model we study here relates to the neglect of spatial interactions. In contrast, a realistic feature missing in the studies mentioned above, and that we explore here, relates to the discrete nature of plants and their intrinsic stochastic behavior.

Due to its relative scarcity, water is the main resource that drives the dynamics of semi-arid ecosystems. It seems natural to model the water densities $\mathbf{x} = (\sigma, \omega)$ as continuous variables governed by an equation which is deterministic when conditioned on the number of plants. More specifically, we will assume that the water dynamics is governed by an equation of the form (\ref{e:general}), but with two continuous variables. Following \cite{Rietkerk-2001,Rietkerk-2002,Rietkerk-2008,Rietkerk-2010} we write $\mathbf{F}=(F_\sigma , F_\omega)$ where
\begin{subequations}\label{e:ns}
\begin{align}
F_\sigma (n, \mathbf{x}) & = R\, -\, \alpha(\rho)\,  \sigma \, , \label{se:ns_surface}\\
F_\omega (n, \mathbf{x}) & = \alpha(\rho)\, \sigma \, -\, \beta(\omega)\, \rho\, -\, r\, \omega\, , \label{se:ns_soil}
\end{align}
\end{subequations}
and where $\rho=\mu n$, $\mu$ being a parameter that characterizes the influence that a single individual P has on the dynamics of the environment. In Eq.~(\ref{e:ns}), $R$ is the average rate of rainfall that increases the amount of surface water $\sigma$, and $r$ is a constant rate that characterizes the loss of soil water which can be due, for instance, to evaporation (see \cite{Rietkerk-2010,Kefi-PhD} for an investigation of a fully deterministic model which considers also loss of surface water). Finally,
\begin{align}\label{e:ab}
\alpha(\rho) = a\: \frac{\rho +k\, W_0}{\rho +k},\hspace{0.5cm}
\beta(\omega) = b\:\frac{\omega}{\omega + k}, 
\end{align}
are saturable rates that describe the infiltration of surface water into the soil and the uptake of soil water by plants, respectively. Here $a,b,k$ and $W_0$ are constants. It is worth mentioning that the infiltration rate is taken to depend on the number of plants in the system, this is in line with  \cite{Rietkerk-2001,Rietkerk-2002,Rietkerk-2008,Rietkerk-2010} and reflects the fact that vegetation typically increases the propensity of surface water being absorbed into the soil. 

We will model plants as discrete entities that follow a stochastic birth-death process. Later on, we shall investigate the impact that demographic noise, due to the discrete nature of plants, has on the behavior of the system. We will assume that a single plant has an average mass $m$ so that the density of biomass per unit area is given by $\rho = m n/A$, where the integer $n$ denotes the number of plants in the system, and where $A$ is the system's total area. It is here important to recall that we are focusing on a well-mixed system and thus we do not consider any spatial structure.  The impact that  plants have on the dynamics of water is via the density $\rho$ (see~Eqs.~(\ref{se:ns_surface}) and (\ref{se:ns_soil})). The effect of the creation or removal of a single plant on the water dynamics is characterized by the parameter $\mu=m/A$, the minimal amount by which the density $\rho$ can change due to a discrete event of the plant dynamics. The birth and death rates for the stochastic plant dynamics are taken to be
\begin{subequations}\label{e:rates}
\begin{align}
\Gamma_b(\omega) &= c\: \beta(\omega),\\
\Gamma_d &= d \, , 
\end{align}
\end{subequations}
respectively, where $d$ and $c$ are constants and $\beta(\omega)$ is defined in (\ref{e:ab}). It should be noticed that the death rate is taken to be constant, whereas the birth rate is assumed to depend on the current amount of soil water. This dependence, and that of $\mathbf{F}$ in Eq.~(\ref{e:ns}) on the density of plants, $\rho = \mu n$, is what couples the deterministic and stochastic dynamics of water and plants, respectively. 

%%%%%%%%%%%%%%%%%%%%%%%%%%
\section{Master equation}\label{s:gme}
%%%%%%%%%%%%%%%%%%%%%%%%%%
We now return to the general development of the formalism, and assume a single discrete variable and a single continuous variable, to avoid cluttering the equations with indices. 

Suppose then that, at time $t$, the system is in state $(n^\prime,x^\prime)$, and denote by $T(n|n^\prime;x^\prime)$ the total rate (i.e., including births, deaths, and any other processes in the model) for the system to make a transition from $n^\prime$ to $n$. The probability that, in a small time interval $\Delta t$, there are no transitions is given by 
\be
p_{\Delta t}^0(x^\prime) = 1-\Delta t \sum_{\ell \neq n^\prime}T(\ell|n^\prime;x^\prime).
\label{p_Delta_t}
\ee
In this case, the system will evolve deterministically according to Eq.~(\ref{e:general}), so that after a time $\Delta t$ it will be in the state $\left(n^\prime, {x}_0\right)$ where
\be \label{e:state1}
{x}_0 = x^\prime + {F}(n^\prime,x^\prime)\Delta t ,
\ee
up to first order in $\Delta t$.

Suppose now that there is a single transition from $n^\prime$ to $n$ which takes place precisely at $\Delta t^\prime < \Delta t$. In this case the system will evolve in a more intricate fashion, to the state $\left(n^\prime, {x}_1\right)$ with
\be\label{e:state2}
{x}_1 = x^\prime + {F}(n^\prime,x^\prime)\Delta t^\prime +{F}(n,x^\prime)\left(\Delta t - \Delta t^\prime\right) ,
\ee
up to first order in $\Delta t$. Notice that the probability to have more than one transition in the interval $\Delta t$ is $\mathcal{O}(\Delta t^2)$. 

Therefore, the probability of a transition during a time $\Delta t$ is composed of two kinds of terms, corresponding to the two possibilities, (\ref{e:state1}) and (\ref{e:state2}), above. Each of these contain a Dirac delta contribution for the piecewise deterministic evolution of ${x}$. Thus the probability of the system being in state $(n,x)$ at time $t+\Delta t$, given it was in state $(n^\prime,x^\prime)$ at time $t$ is given by
\BE
& & \mathcal{P}_{\Delta t}(n,x |{n}^\prime,{x}^\prime) = p_{\Delta t}^{0}({x}^\prime)\, \delta\left(x - {x}_0\right)\delta_{n^\prime,n} \nonumber \\
& & + \Delta t\, T(n|{n}^\prime;x^\prime)\, \delta\left(x - {x}_1\right)\left(1 - \delta_{n^\prime,n}\right),
\EE
with $x_0$ and $x_1$ as defined in Eqs.~(\ref{e:state1}) and (\ref{e:state2}) respectively.
Using Eq.~(\ref{p_Delta_t}) this gives, to first order in $\Delta t$,
\BE
& & \mathcal{P}_{\Delta t}(n,x |{n}^\prime,{x}^\prime) = \delta\left(x - {x}_0\right)\delta_{n^\prime,n} \nonumber \\
& & - \Delta t \sum_{\ell \neq n^\prime}T(\ell|n^\prime;x^\prime)\delta\left(x - x^\prime \right)\delta_{n^\prime,n} \nonumber \\ 
& & + \Delta t\, T(n|{n}^\prime;x^\prime)\, \delta\left(x - x^\prime \right)\left(1 - \delta_{n^\prime,n}\right),
\label{inter}
\EE
where in the last two terms we have replaced $x_0$ and $x_1$ respectively by
$x^\prime$, since these terms are already of order $\Delta t$. We can now use the Chapman-Kolmogorov equation together with Eq.~(\ref{inter}) to obtain a master equation for the evolution of the probability at finite time $t$. The last two
terms on the right-hand side of Eq.~(\ref{inter}) give the standard terms in the master equation, but the first term gives a contribution 
\be
\int dx^\prime \sum_{n^\prime} \delta\left(x - {x}_0\right)\delta_{n^\prime,n}\mathcal{P}(n^\prime,x^\prime,t).
\label{first_term}
\ee
It is worth pointing out that $x_0$ depends on the integration variable $x'$, see Eq.~(\ref{e:state1}). Introducing a test function, and integrating by parts, we find that (e.g. see \cite{Gardiner-book}) this term equals
\be
\mathcal{P}(n,x,t) - \Delta t\,\frac{\partial }{\partial x}\left[ F(n,x) \mathcal{P}(n,x,t) \right],
\label{firstterm}
\ee
to first order in $\Delta t$. This then yields the master equation for the evolution of the probability distribution $\mathcal{P}(n,{x},t)$ for the system to be in a state (${n}, {x})$~\cite{Davis-1984, Faggionato-2009, Faggionato-2010}
\begin{equation}\label{e:master}
\begin{split}
\frac{\partial }{\partial t}\mathcal{P}({n},{x},t) = & -\frac{\partial }{\partial x}\left[{F}({n},{x} ) \mathcal{P}(n,{x},t)\right] \, \\ &+ \sum_{{n}^\prime\neq{n}}\left[{T}(n|{n}^\prime; {x})\mathcal{P}(n^\prime,{x},t)\right. \\ & \left. - {T}(n^\prime|{n} ; {x})\mathcal{P}(n,{x},t)\right].
\end{split}
\end{equation}

Although we have derived the master equation for a PDMP with one discrete and one continuous variable, the generalization to many variables can immediately be seen to be
\begin{equation}\label{e:master_many}
\begin{split}
\frac{\partial }{\partial t}\mathcal{P}(\mathbf{n},\mathbf{x},t) = & -\sum^{C}_{i=1}\frac{\partial }{\partial x_i}\left[{F}_i(\mathbf{n},\mathbf{x} ) \mathcal{P}(\mathbf{n},\mathbf{x},t)\right] \, \\ &+ \sum_{\mathbf{n}^\prime\neq\mathbf{n}}\left[{T}(\mathbf{n}|\mathbf{n}^\prime; \mathbf{x})\mathcal{P}(\mathbf{n}^\prime,\mathbf{x},t)\right. \\ & \left. - {T}(\mathbf{n}^\prime|\mathbf{n} ; \mathbf{x})\mathcal{P}(\mathbf{n},\mathbf{x},t)\right].
\end{split}
\end{equation}
The terms in the second sum are of the standard form one would expect in a master equation for birth-death processes. The first term describes the (piecewise) deterministic evolution of $\mathbf{x}$, and is of the form of a drift term in a standard Fokker-Planck equation for continuous processes. The absence of a term containing second derivatives with respect to components of $\mathbf{x}$ reflects the (piecewise) deterministic nature of the evolution of $\mathbf{x}$.
The master equation fully specifies the dynamics of the stochastic system, but it cannot be solved analytically and it is difficult to solve numerically. Instead, either single trajectories for the underlying process can be simulated by using an algorithm similar to that originally devised by Gillespie~\cite{Gillespie-1976,Gillespie-1977}, see Appendix \ref{a2} for details, or approximation schemes can be applied to the master equation. In the next section we discuss an example of such a scheme.

%%%%%%%%%%%%%%%%%%%%%%%%%%%%%%%% 
\section{Approximate dynamics}\label{s:ad}
%%%%%%%%%%%%%%%%%%%%%%%%%%%%%%%%
The method we will use to analyze the master equation (\ref{e:master_many}) separates out the average behavior from the fluctuations around it. We will use the example of the semi-arid ecosystem in Section \ref{s:ex} to illustrate the idea.
In general, the fluctuations in the discrete variables $n$, are expected to be of order $\sqrt{n}$, and their impact on the deterministic dynamics of order $\sqrt{\mu n}$, where $\mu=m/A$, as defined above, is the minimal change in mass density per unit area induced by a birth or death event. We will thus introduce the change of variables
\be\label{e:n}
\mu n = \rho + \sqrt{\mu}\, \eta,
\ee
where $\rho$ is the deterministic density introduced in Section \ref{s:ex}. To the order that we will be working, $\mu \langle n \rangle = \rho$, where the angle brackets stand for an average with probability density function $\mathcal{P}(n,\mathbf{x},t)$ at time $t$. The stochastic deviation from the deterministic result, given by $\rho$, is described by the term $\sqrt{\mu} \eta$, and so the deterministic limit corresponds to $\mu \to 0$. The physical meaning of this limit is that the area of the system is to be chosen sufficiently large (formally infinite) so that it contains a large number of plants. The discreteness of the birth-death dynamics is then no longer relevant. In addition the system is always assumed to be well-mixed so that any possible spatial structure can be neglected. 

The intrinsic fluctuations in the discrete variables will induce fluctuations in the {continuous} variables $\mathbf{x}$. For this reason we will also carry out the replacement
\be\label{e:x}
\mathbf{x} = \bs{\chi} + \sqrt{\mu}\,\bs{\xi},
\ee
where $\bs{\chi}= \langle \mathbf{x} \rangle$ and $\bs{\xi}$ are the average and {fluctuation} terms in the {continuous} variables, respectively. 

We would like to stress that, up to now, Eqs. (\ref{e:n}) and (\ref{e:x}) are nothing else than a suitable change of variables. We are here assuming that all the stochastic variation is contained in the variables $({\eta}, {\bs{\xi}})$, while the corresponding averages $({\rho},{\bs{\chi}})$, obtained in the limit $\mu\to 0$ as explained above, follow a deterministic dynamics. For this reason we introduce the probability distribution $\Pi ({\eta}, {\bs{\xi}},t)$ of the stochastic variables $({\eta}, {\bs{\xi}})$ alone, in order to separate these two kinds of contributions. Once again we reiterate that the fluctuations are not externally imposed, but are rather intrinsic to the model. Their statistical properties emerge from the model and from the approximations that we carry out, as we discuss in more detail in Section \ref{s:f}. Notice that, in terms of the new variables, a single stochastic transition, $n\to n\pm 1$, corresponds to $\eta\to\eta\pm\sqrt{\mu}$. When $\mu$ is small, the effect of a single transition can, therefore, be conveniently represented by a Taylor expansion in $\sqrt{\mu}$. The detailed calculations follow the lines of \cite{vanKampen-book}, and are summarized briefly in Appendix \ref{a1}.

%%%%%%%%%%%%%%%%%%%%%%%%%%%%%%%%%%
\section{Deterministic limit}\label{s:dl}

%%%%%%%%%%%%%%%%%%%%%%%%%%%%%%%%%%
\begin{figure}[t]
%\vspace{0.5cm}
%\begin{center}
\hspace{-0.5cm}\includegraphics[width=60mm]{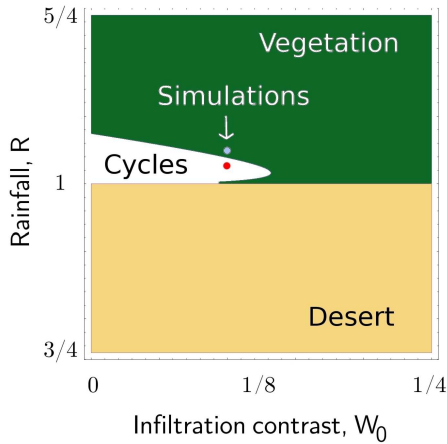}
\caption{(Color online) Phase diagram of the deterministic system, Eqs.~(\ref{e:nsdet}). In the upper region (`Vegetation') the state with vegetation $\rho=\rho^\ast > 0$ is stable under small perturbations, i.e. ${\rm Re}(\lambda_{\rm max}) < 0$. In the small region to the centre left (`Cycles'), there are no stable fixed points, but a numerical integration of Eqs.~(\ref{e:nsdet}) reveals the presence of limit cycles (see discussion in Section \ref{s:f}). In the bottom region (`Desert'), the desert state ($\rho =0$) is stable under small perturbations. The two points marked `Simulations' correspond to $R=1.03$ and $R=1.05$ (with $W_0=0.1$), these are the parameters chosen for the stochastic simulations discussed in Section \ref{s:f}. \label{f:1}}
%\end{center}
\end{figure}

The leading contribution found from the above expansion shows that the average behaviour, described by $\rho$ and $\bs{\chi}$, is given by 
\begin{equation}\label{e:nsdet}
\dv{\rho}{t} = \Phi(\rho,\bs{\chi}),\hspace{1cm}
  \dv{\bs{\chi}}{t} = \mathbf{F}(\rho,\bs{\chi}),
\end{equation}
where
\be\label{e:phi}
\Phi(\rho,\bs{\chi})= \Gamma_b(\bs{\chi})\, \rho - \Gamma_d(\bs{\chi})\, \rho.
\ee
The first equation in (\ref{e:nsdet}) can be found from Appendix \ref{a1} or simply by calculating the average of $\mu n$ from the master equation. The function $\mathbf{F}$ is given by Eq.~(\ref{e:ns}).

We now analyze some of the properties of the limiting deterministic dynamics. The first question we can ask is: are there any fixed points, i.e. do states exist such that
\be\label{e:nsfp}
\dv{\rho}{t}  =  0,\hspace{1cm}\dv{\bs{\chi}}{t}  =  0,
\ee
and, if so, are they stable under small perturbations ?

To study the (linear) stability of these states, it is useful to introduce a small perturbation $\bs{\varepsilon} = (\delta\rho ,\delta\bs{\chi})$ around the fixed point of interest, say $(\rho,\bs\chi)=(\rho^\ast ,\bs{\chi}^\ast)$, which is a solution to (\ref{e:nsfp}) above. We thus write $(\rho,\bs\chi)=(\rho^\ast+\delta\rho,\bs{\chi}^\ast+\delta\bs\chi)$ and expand equations (\ref{e:nsdet}) up to first order in  $\bs\varepsilon$ to obtain
\be\label{e:eps}
\dv{\bs\varepsilon}{t} = \mathcal{J}^\ast\cdot \bs\varepsilon .
\ee
Here $\mathcal{J}^\ast = \mathcal{J}(\rho^\ast,\bs{\chi}^\ast)$ is the $3\times 3$ Jacobian matrix of the system of equations (\ref{e:nsdet}), given by (\ref{e:jac}), evaluated at the fixed point.
The stability of the fixed point is then governed by the eigenvalue with the largest real part, $\lambda_{\rm max}$, of $\mathcal{J}^\ast$. If the real part of this eigenvalue is negative, a small perturbation will die out after a time of order $1/|{\rm Re}(\lambda_{\rm max})|$, otherwise it will grow exponentially fast until non-linearities set in and limit the growth.

We now apply this analysis to Eqs. (\ref{e:nsdet}). Depending on the choice of parameters, there can be either one {out of} two stable fixed points \cite{Rietkerk-2008} or none: the fixed points correspond to either a desert state 
\be
\rho_0 = 0,~\sigma_0 = \frac{R}{a\, W_0},~\omega_0 = \frac{R}{r},
\ee 
or a state with non-zero vegetation 
\be
\rho^\ast = \frac{c\, R}{d}-\frac{r\, c\, k}{c\, b-d},~ \sigma^\ast = \frac{R}{\alpha(\rho^\ast)},~ \omega^\ast = \frac{d\, k}{c\, b-d}.
\ee 
Figure ~\ref{f:1} illustrates these three situations in terms of the parameters $R$ and $W_0$. In the region where there are no stable fixed points a numerical integration of Eqs.~(\ref{e:nsdet}) shows the existence of limit cycles, see Fig.~\ref{f:1} and further discussion in Section \ref{s:f}. These limit cycles were not reported in \cite{Rietkerk-2002}, presumably because the authors worked in a steady-state approximation that neglected perturbations of the surface water density $\sigma$. As our analysis reveals, these perturbations can render the fixed point $\rho=\rho^\ast$ unstable. 

While the range of parameters for which there are cycles in the deterministic approximation appears rather small, it has been observed elsewhere that demographic noise can effectively enlarge such regions \cite{McKane-PRL2005,Boland-2008,Biancalani-2011,Goldenfeld-2011,Biancalani-2010,Goldenfeld-2009}. Noise-induced oscillations can be found in parameter regimes in which a deterministic analysis predicts stable fixed points, i.e. outside the region labelled `Cycles' in Fig. \ref{f:1}.  These so-called quasi-cycles can be characterized by analytical techniques described below, their amplitude is particularly pronounced near the deterministic instability. Our analysis focuses on parameter regimes outside, but near, the region of instability in the deterministic phase diagram of Fig. \ref{f:1}. This is sufficient to illustrate the main point we want to make in this paper: that demographic stochasticity can alter the qualitative behavior of the model relative to that predicted in the deterministic approximation. We also carry out simulations inside the region of parameters in which the deterministic system has limit cycles. Finally we would like to add that when spatial interactions are taken into account, further phases can be found in Fig.~\ref{f:1}. In these phases the model exhibits spatial patterns coexisting with either of the homogeneous states (desert or homogeneous vegetation  respectively), see \cite{Rietkerk-2001,Rietkerk-2002,Rietkerk-2008,Rietkerk-2010} for further details.
%%%%%%%%%%%%%%%%%%%%%%%%%%%%%%%%%%
\begin{figure}[t]
\begin{center}
\includegraphics[angle=-90, width=80mm]{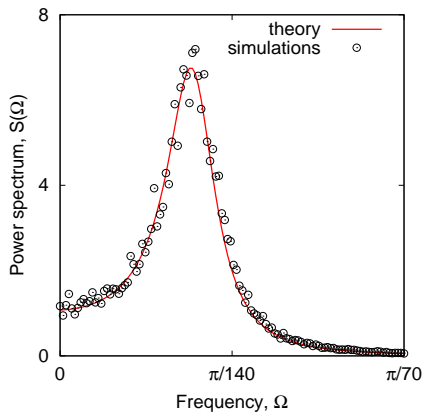}
\caption{(Color online) Power spectrum of the fluctuations of the number of plants in the phase with a stable deterministic fixed point. The solid (red) curve is the analytical prediction of Eqs.~(\ref{e:gps}) and (\ref{e:pps}); symbols show results from simulations of the PDMP, averaged over $100$ realizations, with an initial number of plants $n = 10^5$. The initial biomass $\mu n$, surface water $\sigma$ and soil water $\omega$ are initialized at the fixed point of the deterministic approximation. The parameter values used are given in Appendix \ref{a3} with $R=1.05$ (corresponding to the upper point in Fig. \ref{f:1}). \label{f:2}}
\end{center}
\end{figure}
%%%%%%%%%%%%%%%%%%%%%%%%%%%%%%%%%%
\begin{figure}[t]
\begin{center}
\includegraphics[trim=25mm 15mm 20mm 10mm, width=57mm]{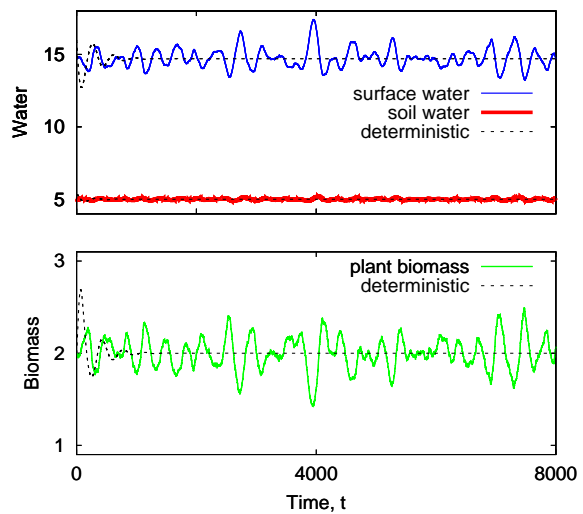}
\caption{(Color online) Noise-induced oscillations in the piecewise deterministic model for semi-arid ecosystems. The continuous (colored) curves show the results of simulations with initial number of plants $n = 10^5$; biomass $\mu\, n$, surface water $\sigma$ and soil water $\omega$ are initialized at the fixed point of the deterministic approximation (dashed lines). From top to bottom the continuous lines are: surface water (blue), soil water (red thick line) and biomass density (green). The (black) dashed lines show the evolution of the corresponding deterministic approximation initialized slightly away of the fixed point to show the oscillatory convergence. This feature is what couples with the noise to induce quasi-cycles in the full model. The parameter values used are those given in Appendix \ref{a3} with $R=1.05$ (upper point in Fig. \ref{f:1}). \label{f:3}}
\end{center}
\end{figure}
%%%%%%%%%%%%%%%%%%%%%%%%%%%%%%%%%%

%%%%%%%%%%%%%%%%%%%%%%%%%%%%%%%%%%
\section{Fluctuations}\label{s:f}
%%%%%%%%%%%%%%%%%%%%%%%%%%%%%%%%%%
Using an expansion in the model parameter $\mu$ (effectively a system-size or small-noise expansion) it is possible to derive a set of coupled linear Langevin equations that approximate the stochastic dynamics close to a stable fixed point $(\rho^\ast,\bs{\chi}^\ast)$. The expansion method goes back to \cite{vanKampen-book} and is standard by now, see \cite{McKane-PRL2005,Boland-2008,Boland-2009,Rogers-2012,Biancalani-2011,Goldenfeld-2011,Scott-2011,Biancalani-2010,Goldenfeld-2009,Lugo-2008}, so that we do not give full details here. A brief summary can be found in Appendix \ref{a1}, including the derivation of the Langevin equation (\ref{e:langevin}) which appears in the sub-leading order of the van Kampen expansion. The stability of the deterministic fixed point is required in order for the stochastic dynamics to remain close to the deterministic attractor, so that the linear approximation remains valid.  

As detailed in Appendix \ref{a1}, carrying out a Fourier transform of this linear Langevin dynamics with respect to time allows one to calculate the power spectra of fluctuations. In particular, we are interested in the power spectrum, $S(\Omega)=\left\langle\left|\widetilde{\eta}(\Omega)\right|^2 \right\rangle$, of the fluctuations of the plant density about the deterministic fixed point, $\rho^\ast >0$. The quantity $\widetilde{\eta}(\Omega)$ is the Fourier transform (with respect to time) of the fluctuations of the plant density, and as before $\avg{\cdots}$ denotes an average over realizations of the stochastic dynamics. Within the van-Kampen expansion analytical results can be derived for the power spectrum, $S(\Omega)$, details can be found in the Appendix, see in particular Eqs.~(\ref{e:gps}) and (\ref{e:pps}). It is worth pointing out that although we here focus on fluctuations of the plant density, similar results can be derived for the fluctuations of the soil and surface water densities.
%%%%%%%%%%%%%%%%%%%%%%%%%%%%%%%%%%
\begin{figure}[t]
\begin{center}
\includegraphics[trim=25mm 5mm 20mm 10mm, width=60mm]{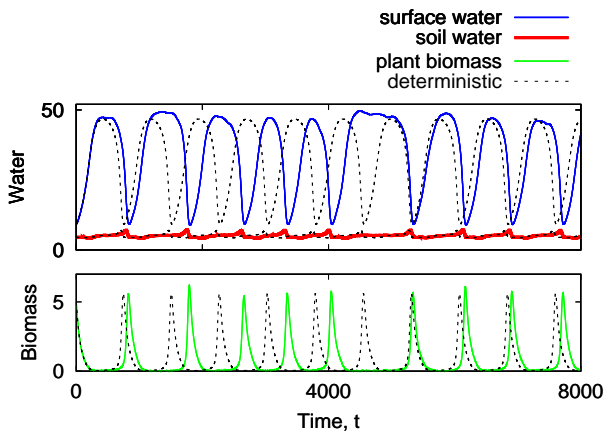}
\caption{(Color online) Effect of demographic noise on the limit cycle of the deterministic model. The continuous curves are results from simulations of the PDMP, with initial number of plants $n = 10^5$; biomass $\mu\, n$, surface water $\sigma$ and soil water $\omega$ are initialized on the limit cycle. Dashed lines represent the deterministic limit cycle. From top to bottom the continuous lines are: surface water (blue), soil water (red thick line) and biomass density (green). The rainfall parameter is $R=1.03$ (corresponding to the lower point indicated in Fig. \ref{f:1}), remaining parameter values can be found in Appendix \ref{a3}. \label{f:4}}
\end{center}
\end{figure}

%%%%%%%%%%%%%%%%%%%%%%%%%%%%%%%%%%
\begin{figure}[t!]
\begin{center}
\includegraphics[trim=10mm 35mm 30mm 35mm, width=60mm]{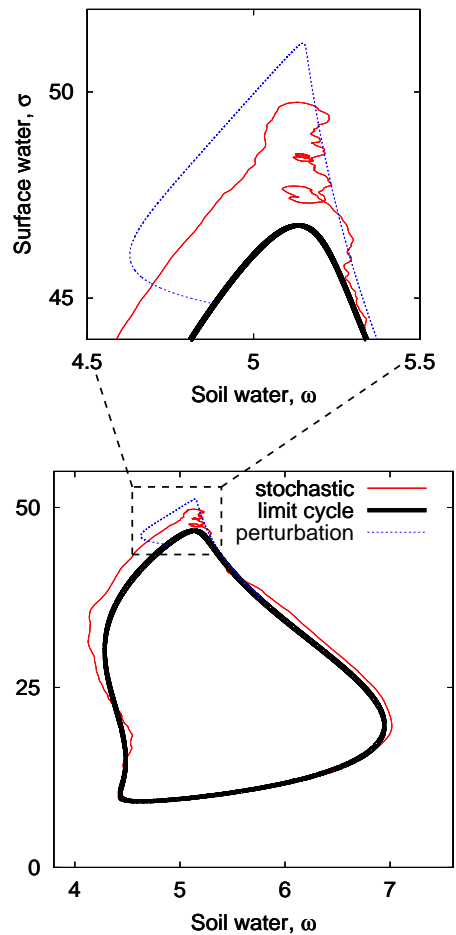}
\caption{(Color online) Projection on the plane of surface and soil water variables of the limit cycle (black thick line) in Fig.~\ref{f:4}. The continuous `wiggly' (red) curve, near to the limit cycle, corresponds to the part of the stochastic trajectory in Fig.~\ref{f:4} between times $4200$ and $5500$ (wide valley in biomass and wide plateau in surface water). It appears that the region where the noise has stronger effects is in the top of the plot, which indeed corresponds to biomass close to zero (so number of plants relatively small). For reference, the dotted (blue) curve at the top corresponds to the deterministic approximation initialized at the limit cycle, except for the biomass which is slightly displaced from the limit cycle ($\rho\approx 0.003$ rather than $0.05$). We can see that the perturbation initially grows away from the cycle and later returns to it. 
\label{f:5} }
\end{center}
\end{figure}
%%%%%%%%%%%%%%%%%%%%%%%%%%%%%%%%%%
We have also carried out numerical simulations using a modification of the Gillespie algorithm \cite{Gillespie-1976,Gillespie-1977}, as described in Appendix \ref{a2}. In Fig. \ref{f:2} we test {the} analytical results {we have just discussed against direct simulations of the PDMP. The value of the parameter $R$ was set to $1.05$. Good agreement between the theoretical predictions and the numerical simulations is found. The figure demonstrates the existence of coherent quasi-cycles, driven by intrinsic noise, in a region of parameter space where the deterministic approximation predicts a stable fixed point, i.e.~where one has ${\rm Re}(\lambda_{\rm max})<0$. As seen in Fig.~\ref{f:2},  the power spectrum, $S(\Omega)$, shows a maximum at a characteristic non-zero frequency, confirming noise-induced oscillations~\cite{McKane-PRL2005}.   An inspection of a single trajectory, as shown in Fig.~\ref{f:3}, shows that these quasi-cycles can be also detected by eye in the time domain.

In Fig.~\ref{f:4} we show similar simulations for $R=1.03$. The deterministic system then no longer has a stable fixed point, but instead it has a limit cycle. The effect of demographic noise on the limit cycle is to introduce a stochastic modulation of the period and amplitude of the cycle. These effects can be studied analytically by separating directions perpendicular to the limit cycle from longitudinal modes in a co-moving Frenet frame. This is discussed in more detail for a different model system in \cite{Boland-2009, Boland-2008}. For the present system, however, we will not investigate this further analytically, but only discuss a few qualitative features. 

Figure \ref{f:5} shows a projection of the limit cycle (black thick continuous line), onto the plane spanned by the surface water and soil water variables. In the same figure we also plot a part of the stochastic trajectory (red thin continuous line) shown in Fig.~\ref{f:4}, specifically Fig. \ref{f:5} shows the segment of the stochastic trajectory between time $4200$ and time $5500$. This segment corresponds to the wide plateau of surface water concentration $\sigma\approx 50$ seen in the upper panel of Fig. \ref{f:4} (upper thin solid line). In this segment the plant biomass, a measure for the number of individuals in the system, remains relatively small, see the solid line in the lower panel of Fig. \ref{f:4}. In this regime of small numbers of individuals the effects of demographic noise are stronger than in periods in which there are more individuals present in the system. This is seen in Fig.~\ref{f:5}; the stochastic trajectory deviates substantially from the deterministic trajectory in the upper part of the limit cycle when the plant biomass is small, but it follows the deterministic cycle more closely when the number of plants is large (lower part of the limit cycle). As shown in \cite{Boland-2009, Boland-2008} the overall effect of demographic noise on a limit cycle is determined by two different factors: one is the relative magnitude of discretization effects, these are small for large populations, but more relevant for small populations as just discussed. Secondly, the local stability of the deterministic trajectory plays an important role as well. For highly stable deterministic trajectories the amplification effect that demographic noise undergoes is relatively small. If the deterministic attractor is only weakly stable, then the amplification factor can be significant. In the case of a fixed point, stability of the deterministic system is characterized by the eigenvalue $\lambda_{\rm max}$. For limit cycles the situation is more complicated, stability is then governed by the relevant Floquet exponents \cite{Boland-2009, Boland-2008}, and crucially local stability can vary along the limit cycle. In our system we observe that stability in the upper part of the limit cycle is only relatively weak. To illustrate this we have initialized the deterministic system at a point near the limit cycle trajectory, but with a slight displacement in the biomass ($\rho\approx 0.003$ instead of $\rho\approx 0.05$). No perturbation is applied to the water variables $\sigma$ and $\omega$. As shown in Fig. \ref{f:5} (blue dotted line) we find that the perturbation initially appears to grow, but ultimately returns back to the limit cycle. This relatively weak stability of the upper part of the deterministic limit cycle may enhance the effects of demographic stochasticity. 

%%%%%%%%%%%%%%%%%%%%%%%%%%%
\section{Conclusions}\label{s:c}
%%%%%%%%%%%%%%%%%%%%%%%%%%%

In this paper we have studied the effects of demographic noise in the framework of piecewise deterministic Markov processes. Specifically we have investigated a class of hybrid systems, composed of $C$ continuous degrees of freedom, and $D$ discrete ones. Such systems are of interest for example in the context of ecology, where the continuous degrees of freedom can represent variables of the external environment such as water or light. The discrete degrees of freedom could then represent the individual-based dynamics of a population of plants or animals. While the population of individuals follows a standard Markovian birth-death process and while the continuous degrees of freedom are governed by ordinary differential equations both aspects are not independent. Non-trivial behavior arises through the coupling of both types of dynamics, for example the birth and death rates will depend on the availability of environmental resources, while these in turn depend on the state of the discrete population. 

We have shown how to apply techniques of nonequilibrium physics to analyze the effects of demographic stochasticity on such systems. In particular, we have shown how to use a system-size expansion to obtain a deterministic approximation in the limit of infinite populations, and to derive an effective Gaussian description of fluctuations about the deterministic behavior in the limit of large, but finite populations. We have applied these concepts to a stylized non-spatial model of semi-arid ecosystems and find that demographic noise can induce persistent coherent stochastic oscillations in parameter regimes in which a deterministic description would predict a stable fixed point. These quasi-cycles can be characterized analytically by deriving closed-form expressions for their power spectra within the small-noise approximation, in good agreement with simulations. This extends existing work on the qualitative features (here quasi-cycles) generated by amplified demographic noise, see \cite{McKane-PRL2005,Boland-2008,Boland-2009,Rogers-2012,Biancalani-2011,Goldenfeld-2011,Scott-2011,Biancalani-2010,Goldenfeld-2009,Lugo-2008}. In numerical simulations we have also investigated the effects of intrinsic noise on deterministic limit cycles, similar to the observations made in \cite{Boland-2009,Boland-2008} we observe that stochasticity can lead to a longitudinal phase-drift along the limit cycle, coupled with transverse fluctuations.

While our present approach is limited to non-spatial hybrid systems, work in progress will extend these results to piecewise deterministic models with spatial interactions. This is particularly interesting for more realistic models of semi-arid ecosystems, in which pattern forming mechanisms are known to be important~\cite{Meron-2011,Rietkerk-2010,Rietkerk-2008,Meron-2004,Rietkerk-2002,Meron-2001}.

\begin{acknowledgements}
We would like to thank Max Rietkerk, Mara Baudena and Maarten Eppinga for insightful discussions. This work was supported by EPSRC under grant RESINEE (reference EP/I019200/1).  TG would like to thank Research Councils UK for support (RCUK reference EP/E500048/1).
\end{acknowledgements}

%%%%%%%%%%%%%%%%%%%%%%%%%%%
%%%%%%%%%%%%%%%%%%%%%%%%%%%
\appendix
%%%%%%%%%%%%%%%%%%%%%%%%%%%
%%%%%%%%%%%%%%%%%%%%%%%%%%%
\section{System-size expansion}\label{a1}
%%%%%%%%%%%%%%%%%%%%%%%%%%%
In this Appendix we briefly describe some of the intermediate steps of the system-size expansion for PDMP in the particular case of the model system which we have focused on in this paper. Introducing the step-operators~\cite{vanKampen-book}
\be
\epsilon^\pm f(n) = f(n\pm 1),
\ee
acting on functions $f({n})$, the master equation (\ref{e:master_many}) can be written as
\BE\label{e:master2}
 \frac{\partial }{\partial t}{\cal P}(n,{\bx},t) &= & -\nabla_{{\mathbf{x}}}\cdot \left[{\bF}(n,\bx ) {\cal P}(n,\bx,t)\right] \nonumber \\ 
 &&+ (\epsilon^--1)\left[T_b(n+1|n,\bx){\cal P}(n,\bx,t)\right]\nonumber \\ 
&& + (\epsilon^+-1) \left[T_d(n-1|n,\bx) {\cal P}(n,\bx,t)\right],\nonumber \\
\EE 
where $\bx=(\sigma,\omega)$ for our model system.
 
In order to make the change of variables described in Eqs.~(\ref{e:n}) and (\ref{e:x}), we introduce $\Pi({\eta},{\bxi},t) = {\cal P}(n,\bx,t)$, and use the relation~\cite{vanKampen-book}
\be\label{e:dt}
\begin{split}
\frac{\partial}{\partial t}{\cal P}(n,\bx,t) = & \,\frac{\partial}{\partial t}\Pi({\eta},{\bxi},t) -\frac{1}{\sqrt{\mu}}\, \dot{{\rho}}\,  \partial_{{\eta}}\Pi({\eta},{\bxi},t) \\ & -\frac{1}{\sqrt{\mu}}\, \dot{{\bchi}}\cdot \nabla_{{\bxi}}\Pi({\eta},{\bxi},t).
\end{split}
\ee
Next the step-operators can be conveniently represented by a Taylor expansion in $\sqrt{\mu}$
\be
\epsilon^\pm  = 1\pm \sqrt{\mu}\,\frac{\partial}{\partial\eta} + \frac{\mu}{2}\frac{\partial^2}{\partial\eta^2}+\mathcal{O}(\mu^{3/2}),
\ee
and similarly, the transition rates can be expanded up to first order in $\sqrt{\mu}$. For instance
\be
\begin{split}
T_b(n+1|n;\omega)=&\frac{1}{\mu}(\rho + \sqrt{\mu}\,\eta)\Gamma_b\left(\bar{\omega}+\sqrt{\mu}\,\xi_\omega\right)\\ \approx& \frac{1}{\mu}\left(\rho + \sqrt{\mu}\,\eta\right)\left[\bar{\Gamma}_b+\sqrt{\mu}\,\xi_\omega\,\bar{\Gamma}_b^{\prime}\right],
\end{split}
\ee
where $\bar\omega$ stands for the concentration of soil water in the deterministic system, and where $\bar{\Gamma}_b = \Gamma_b(\bar{\omega})$ and $\bar{\Gamma}_b^{\prime}=d\bar\Gamma/d\bar{\omega}$, evaluated at the deterministic value $\bar\omega$.

Inserting the various ingredients into Eq.~(\ref{e:master2}) and picking out the term of order $\mu^{-1/2}$ one finds
\be
\dot{{\rho}}\, \partial_{{\eta}}\Pi + \dot{{\bchi}}\cdot \nabla_{{\bxi}}\Pi = {\Phi}\,\partial_{{\eta}}\Pi + {\bF}\cdot\nabla_{{\bxi}}\Pi,
\ee
with ${\Phi}=\Phi(\rho,\boldsymbol{\chi})$. 
This equation is satisfied if $\rho$ and $\boldsymbol{\chi}$ fulfill the deterministic evolution equations (\ref{e:nsdet}).

Analyzing the next-to-leading term in the expansion (i.e. the term of order $\mu^0$) we obtain a Fokker-Planck equation  
\be\label{e:fokker}
{\partial_t} \Pi= \left\{-\nabla_{\bzeta}\cdot\left[\mathcal{J}\cdot{\bzeta}\,\Pi\right]+\frac{1}{2}\partial_{\eta}^2\left[B\Pi \right]\right\}.
\ee
Here we have used the notation $\bzeta \equiv (\eta,\bxi)=(\eta,\xi_\sigma,\xi_\omega)$. The quantity
\be\label{e:jac}
\mathcal{J}= \begin{pmatrix}
\partial_{\rho} \Phi(\rho,\bs\chi) & \nabla_{\bchi}\Phi(\rho, \bs\chi) \\ 
\partial_{\rho} \bF(\rho, \bs\chi) & \nabla_{\bchi}\bF(\rho,\bs \chi)
\end{pmatrix},
\ee
is the ($3\times 3$) Jacobian matrix of the deterministic model, and
\be\label{e:fokker_coeff}
B = \left[ d + \Gamma_b(\bar{\omega})\right] \, \rho 
\ee
is a  diffusion coefficient that describes the strength of the noise.

The Fokker-Planck equation (\ref{e:fokker}) is equivalent to a linear Langevin equation
\be\label{e:langevin}
\dv{\bzeta}{t} = \mathcal{J}\cdot{\bzeta} + \bnu(t) ,
\ee
with a noise term given by $\bnu \equiv (\vartheta, 0, 0)$ and
\be
\langle\vartheta(t)\vartheta(t^\prime)\rangle = B\delta(t-t^\prime).
\ee
We notice that the second and third components of $\bnu$ vanish, all sources of the demographic stochasticity are now contained in the first component of $\bnu$, and given by the noise variable $\vartheta(t)$. This reflects the fact that only the variable $n$ is subject to demographic discretization, whereas $\bx=(\sigma,\omega)$ is continuous. However, given that the three variables of the linearized dynamics are coupled (i.e. the Jacobian matrix ${\cal J}$ will not generally be diagonal), this does not mean that $\xi_\sigma$ and $\xi_\omega$ will vanish. 

It is also interesting to note that the Jacobian matrix ${\cal J}$ and the variance $B$ of the noise variable $\eta$ can be expressed purely in terms of quantities derived from the deterministic dynamics. During the transient phase of the dynamics ${\cal J}$ and $B$ will be time-dependent, but in order to proceed analytically we will focus on the effect of small fluctuations about a fixed point $(\rho,\bs\chi) = (\rho^\ast, \bs{\chi}^\ast)$. The quantities ${\cal J}$ and $B$ are then constants. 

Carrying out a Fourier transform of the above Langevin equation we find
\be\label{e:langevin_full_fourier}
i\, \Omega\,\widetilde{\bzeta}(\Omega) = {\mathcal{J}}^{\ast}\cdot\,\widetilde{\bzeta}(\Omega) + \widetilde{\bnu}(\Omega) ,
\ee 
with $\widetilde{\bnu} =( \widetilde{\vartheta}, 0, 0)$ and
\be\label{e:noise}
{\left\langle\widetilde{\vartheta}(\Omega)\,\widetilde{\vartheta}(\Omega^\prime)\right\rangle} = 2\pi\,\delta(\Omega+\Omega^\prime)B^\ast,
\ee
where the asterisk indicates that the corresponding objects have been evaluated at the deterministic fixed point. The algebraic equation (\ref{e:langevin_full_fourier}) can be solved for $\widetilde\bzeta(\Omega)$ to find
\be\label{e:fourier_solution}
\widetilde{\bzeta}(\Omega) = \left[i\, \Omega\id - {\mathcal{J}}^\ast\right]^{-1} \widetilde{\bnu}(\Omega) ,
\ee 
when the inverse exists. This allows us to compute the power spectrum of the fluctuations which, by applying Cramer's rule, can be written as
\be\label{e:gps}
\left\langle\left|\widetilde{\zeta}_s(\Omega)\right|^2\right\rangle = \frac{B}{\left|\det \left(i \Omega - {\mathcal{J}}^\ast \right)\right|^2}\left|\mathcal{M}_{1 s}(\Omega)\right|^2,
\ee
where $s=1,2,3$, with $\widetilde{\bzeta} = (\widetilde{\eta},\widetilde{\xi}_{\sigma}, \widetilde{\xi}_{\omega})$, and $\mathcal{M}_{1\,s}$ is given by the $(1, s)$ minor of the matrix $i\, \Omega \id- {\mathcal{J}}^\ast$, i.e. the determinant of the matrix obtained after deleting its row 1 and column $s$. Specifically, we are interested in the spectral decomposition of the fluctuations in the number of plants (i.e. $s=1$), for which
\be\label{e:pps}
\left|\mathcal{M}_{11}(\Omega)\right|^2 = \left(\Omega^2+\left|{\mathcal{J}}^\ast_{22}\right|^2\right)\left(\Omega^2+\left|{\mathcal{J}}^\ast_{33}\right|^2\right).
\ee

%%%%%%%%%%%%%%%%%%%%%%%%%%%%%%%%%%%%
\section{Simulation algorithm for piecewise deterministic Markov processes}\label{a2}
%%%%%%%%%%%%%%%%%%%%%%%%%%%%%%%%%%%%
In order to generate trajectories of the piecewise deterministic Markov process defined by Eqs. (\ref{e:general}) and (\ref{e:general_rules}), i.e. to sample solutions of the master equation (\ref{e:master_many}), we use a modification of the celebrated Gillespie algorithm \cite{Gillespie-1976, Gillespie-1977}. Here we describe one iteration of the simulation in the case of a PDMP with one discrete and one continuous variable. The generalization to many discrete and continuous variables is straightforward. Starting at time $t$ with the system in state $(n,x(t))$,
to carry out the next simulation step, we first draw a time increment $\tau$ from the distribution defined by (see e.g. \cite{Zeiser-2008,Faggionato-2009})
\be\label{e:taudist}
\mbox{Prob}(\tau\leq\theta)=1-\exp\left(-\int_t^{t+\theta} T\left(n; x(s)\right)\mathrm{d} s\right).
\ee
Here we have written
\be
T(n;x) = T_b(n+1|n; x)+T_d(n-1|n; x),
\ee
which is the total rate for a reaction to occur if the system is in state $(n,x)$. The quantity $x(s)$ is given by the solution of Eq. (\ref{e:general}) with initial condition $x(t)$, $s\ge t$. Generating time increments from the distribution above is not entirely straightforward, we will describe the details below.

Once the increment $\tau$ has been generated, i.e. the time of the next event has been determined, we need to decide whether this event is a birth or death event. This is carried out following the standard procedure of the Gillespie algorithm using the appropriate probabilities given by the reaction rates at time $t+\tau$ i.e.,~a birth event will occur with probability $T_b(n+1|n,x_\tau)/T(n,x_\tau)$, with $x_\tau = x(t+\tau)$, and a death event with the complementary probability $T_d(n+1|n,x_\tau)/T(n,x_\tau)$. The event is then carried out in the simulation, increasing or decreasing the number of plants $n$ by one. So the density is $n(t)=n$ prior to the simulation step, and $n(t+\tau)=n\pm 1$ afterwards. Given that the actual birth or death event occurs at $t+\tau$ we have $n(t')=n(t)$ for all $t'\in[t,t+\tau)$. Likewise, the state of the continuous variable at time $t+\tau$ is given by the solution $x(t+\tau)$ of the differential equation $\dot x=F(n,x)$, with initial condition $x(t)$; here the first argument of $F$ is given by the number of individuals $n$ at the beginning of the simulation step.

It now remains to explain in more detail how we sample the probability distribution defined by Eq. (\ref{e:taudist}), i.e. how the time increments $\tau$ are generated. This is done by first drawing a number $r$ uniformly at random in $(0,1]$ and then solving
\be\label{e:lnr}
\int_t^{t+\tau} T(n; x(s))\mathrm{d} s = \ln (1/r)
\ee
for $\tau$. We note, though, that the dependence of the left-hand side on the variable $\tau$ is intricate, so we need to use a numerical approximation scheme to evaluate the integral while at the same time determining its upper limit $\tau$ self-consistently. Suppose that $\Delta t$ is a time step suitable for the numerical integration of the differential equations $\dot x=F(n,x)$, for every $n$. We introduce the notation $x_\ell=x(t+\ell\Delta t)$ and $T_\ell=T(n,x_\ell)$. The algorithm then proceeds as follows:
\begin{enumerate}
\item[1.] In the first instance simply approximate the integral by $T_0\tau$. Substitute this into Eq. (\ref{e:lnr}) and determine the resulting value for $\tau$. If $\tau$ is found to be smaller than $\Delta t$, use the resulting value as an approximation for the time increment, and carry out the Gillespie step.
\item[2.] If $\tau$ is found to be greater than $\Delta t$ we can improve the approximation of the integral used in step 1: approximate the integral by $T_0\Delta t+(\tau-\Delta t)T_1$. Substitute this into Eq. (\ref{e:lnr}) and determine $\tau$. If $\tau$ is found to be smaller than $2\Delta t$, use this as an approximation for the time increment.
\item[3.] Sequentially continue this scheme to the case of an increasing number of $M=2,3,...$ discretization steps, i.e. approximate the integral by $\Delta t\sum_{\ell=0}^{M-1}T_\ell+(\tau-M\Delta t)T_M$ and solve Eq. (\ref{e:lnr}) for $\tau$. If $\tau$ is found to be smaller than $(M+1)\Delta t$, terminate, otherwise increase $M$ by one and iterate.
\end{enumerate}
While this algorithm may appear quite tedious and costly in terms of CPU time, we typically find in our simulations that the scheme terminates after a few steps when $\mu$ is small; see also  \cite{Zeiser-2008} for a different approach.

%%%%%%%%%%%%%%%%%%%%%%%%%%%%%%%%%%%%
\section{Parameter values}\label{a3}
%%%%%%%%%%%%%%%%%%%%%%%%%%%%%%%%%%%%
Except when specified otherwise, the parameter values used in the simulation are chosen in analogy with \cite{Rietkerk-2010, Rietkerk-2008, Rietkerk-2002} and are $a=0.2, b=0.05, c=10, d=0.25, r=0.2, k=5$ and $W_0 = 0.1$.

%\bibliography{semiarid.bib}

\end{document}